\documentclass[aps,prl,twocolumn,superscriptaddress,calc,epsfig,10pt]{revtex4-1}
\usepackage{graphicx}
\usepackage{graphics}
\usepackage{amssymb,amsmath}

\usepackage{wasysym}

\usepackage{color}

\newcommand{\mitCUAaddress}{Department of Physics, MIT-Harvard Center for Ultracold Atoms, and Research Laboratory of Electronics, MIT, Cambridge, Massachusetts 02139, USA}

\newcommand{\ket}{\right\rangle}

\newcommand{\bra}{\langle}

\begin{document}
\title{A Quantum Gas Microscope for Fermionic Atoms}

\author{Lawrence W. Cheuk}
\author{Matthew A. Nichols}
\author{Melih Okan}
\author{ Thomas Gersdorf}
\author{Vinay V. Ramasesh}
\author{Waseem S. Bakr}
\author{Thomas Lompe}
\author{Martin W. Zwierlein}

\affiliation{\mitCUAaddress}

\begin{abstract}
Strongly interacting fermions define the properties of complex matter throughout nature, from atomic nuclei and modern solid state materials to neutron stars. Ultracold atomic Fermi gases have emerged as a pristine platform for the study of many-fermion systems. Here we realize a quantum gas microscope for fermionic $^{40}$K atoms trapped in an optical lattice, which allows one to probe strongly correlated fermions at the single atom level. We combine 3D Raman sideband cooling with high-resolution optics to simultaneously cool and image individual atoms with single lattice site resolution at a detection fidelity above $95\%$. The imaging process leaves the atoms predominantly in the 3D motional ground state of their respective lattice sites, inviting the implementation of a Maxwell's demon to assemble low-entropy many-body states. Single-site resolved imaging of fermions enables the direct observation of magnetic order, time resolved measurements of the spread of particle correlations, and the detection of many-fermion entanglement.
\end{abstract}
\maketitle

\begin{figure}
\centering
\includegraphics[scale=1.0]{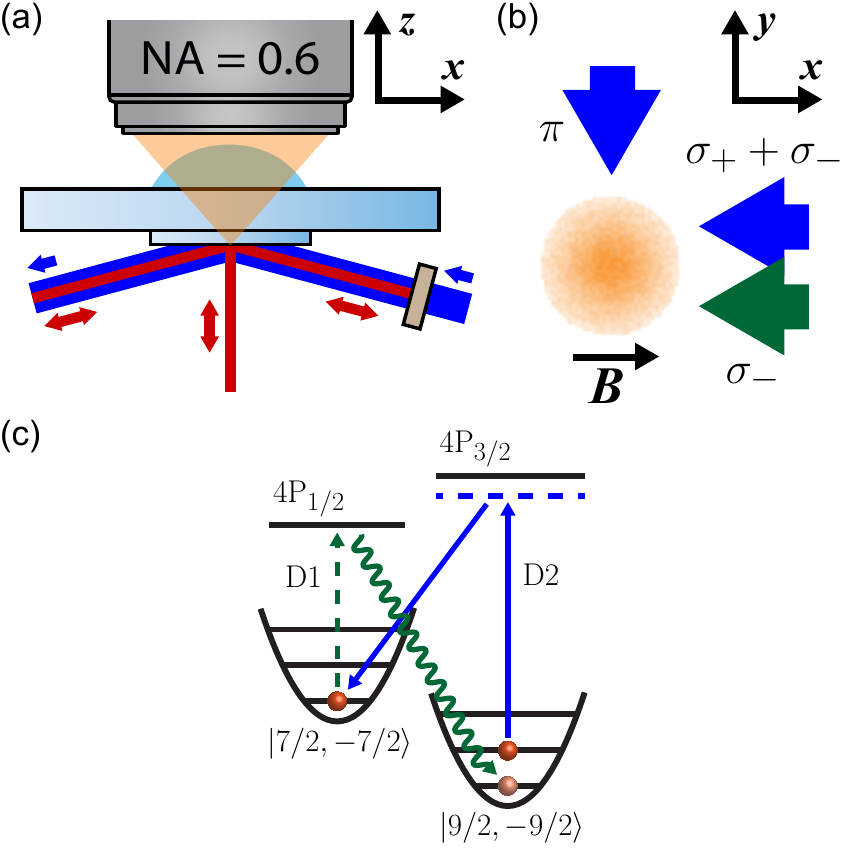}
\caption{(a) High-resolution imaging setup. A solid immersion lens is formed by a spherical cap and a super-polished substrate contacted on either side of the vacuum window. Using an objective with NA = 0.60, the system achieves an effective NA = 0.87. The substrate reflects $1064\,\rm nm$ light while transmitting D1 and D2 light of $^{40}$K. 
The lattice beams are shown in red; the optical pumping and $x$-Raman beams are shown in blue. (b) Top view of Raman beams (blue) and optical pumping beam (green). (c) Raman cooling scheme. The Raman beams detuned near the D2 line (solid blue) drive vibration-lowering transitions. The optical $F$-pumping beam (dashed green) is tuned to the D1 line. Not shown is the $m_F$-pumping beam.}
\label{fig1}
\end{figure}

The collective behavior of fermionic particles governs the structure of the elements, the workings of high-temperature superconductors and colossal magneto-resistance materials, and the properties of nuclear matter. Yet our understanding of strongly interacting Fermi systems is limited, due in part to the antisymmetry requirement on the many-fermion wavefunction and the resulting ``fermion sign problem''. In recent years, ultracold atomic quantum gases have enabled quantitative experimental tests of theories of strongly interacting fermions~\cite{ingu08varenna, bloc08review, Zwerger2011BECBCS,Zwierlein2014NovelSuperfluids}. In particular, fermions trapped in optical lattices can directly simulate the physics of electrons in a crystalline solid, shedding light on novel physical phenomena in materials with strong electron correlations. A major effort is devoted to the realization of the Fermi-Hubbard model at low entropies, believed to capture the essential aspects of high-$T_c$ superconductivity~\cite{Esslinger2010FermiHubbard,chin06,jord08,schn08,Greif2013Magnetism,Imriska2014Hubbard,Hart2014FermiHubbard}.
For bosonic atoms, a new set of experimental probes ideally suited for the observation of magnetic order and correlations has become available with the advent of quantum gas microscopes~\cite{bakr2009microscope,sherson2010microscope,miranda2014}, enabling high-resolution imaging of Hubbard-type lattice systems at the single atom level. They allowed the direct observation of spatial structures and ordering in the Bose-Hubbard model~\cite{bakr2010MottInsulator,sherson2010microscope} and of the intricate correlations and dynamics in these systems~\cite{Endres2011stringorder,Cheneau2012Lightcone}.
A longstanding goal has been to realize such a quantum gas microscope for fermionic atoms. This would enable the direct probing and control at the single lattice site level of strongly correlated fermion systems, in particular the Fermi-Hubbard model, in regimes that cannot be described by current theories. These prospects have sparked significant experimental effort to realize site-resolved, high-fidelity imaging of ultracold fermions, but this goal has so far remained elusive.

In the present work, we realize a quantum gas microscope for fermionic $^{40}$K atoms by combining 3D Raman sideband cooling with a high resolution imaging system. The imaging setup incorporates a hemispherical solid immersion lens optically contacted to the vacuum window (Fig.~1(a)). In combination with a microscope objective with numerical aperture (NA) of 0.60, the system achieves an enhanced NA of 0.87 while eliminating aberrations that would arise from a planar vacuum window.
In order to keep the atoms localized while performing fluorescence imaging, one must simultaneously cool them in order to mitigate the heating from spontaneously emitted imaging photons. Previous microscope experiments in Hubbard-type lattices~\cite{bakr2009microscope,sherson2010microscope,miranda2014} cool via optical molasses. In contrast, we employ 3D Raman sideband cooling~\cite{Monr95RSC,Hamann98, Vuletic1998RSC,Kerm00,Han00,nelson07,kaufman2014,patil2014,thompson14}, in which Raman transitions on vibration-lowering sidebands are combined with optical pumping to provide cooling.
Our method therefore not only achieves site-resolved imaging, but also leaves a large fraction of the atoms ($72(3)\%$) in the 3D motional ground state of each lattice site. This opens up prospects for the preparation of low entropy many-body states, by measuring the atoms' initial positions and rearranging them into the desired configuration \cite{weiss2004}.

Raman sideband cooling has previously been used to cool  $^{87}$Rb and $^{133}$Cs atoms in lattices and in optical tweezers to large ground state populations ~\cite{Hamann98, Vuletic1998RSC, Kerm00, Han00, nelson07,kaufman2014, patil2014, thompson14}. 
Here, we realize continuous Raman sideband cooling of $^{40}$K using two states from the ground hyperfine manifolds, $\left|a\ket=\left|F=9/2, m_F=-9/2\ket$ and $\left|b\ket= \left|7/2, -7/2\ket$, which form an approximate two-level system. To make $\left|a\ket$ and $\left|b\ket$ non-degenerate with other hyperfine states, and to provide a quantization axis for optical pumping, we apply a bias field of $4.2\,\rm G$ along the $x$ direction (Fig.~1(b)). A pair of Raman beams collinear with $x$ and $y$ lattice beams, but not retro-reflected, drives vibration-lowering Raman transitions from $\left|a\ket$ to $\left|b\ket$ (Fig.~1(c)). The Raman lasers are detuned $-41\,\rm GHz$ from the D2 transition. The optical pumping is performed on the D1 transition, $3\,\rm nm$ away from the D2 line, allowing us to filter out stray Raman light while transmitting atomic fluorescence. By collecting the photons that are spontaneously scattered during this optical pumping process, we can image the atoms without using additional resonant light.

To prepare a cold cloud of fermionic atoms under the microscope, $^{40}$K is first sympathetically cooled with $^{23}$Na in a plugged magnetic quadrupole trap~\cite{Park:2012}, centered $\sim9\,\rm mm$ below a super-polished substrate that forms the bottom of the solid immersion lens. After removal of $^{23}$Na, the cloud of $\sim1$ million $^{40}$K atoms is magnetically transported to the substrate and trapped in a vertical lattice formed by a $1064\,\rm nm$ laser beam reflected off the substrate at an angle of $5.9^\circ$. A single layer $7.8\,\mu$m from the surface is selected using a radiofrequency sweep in a vertical magnetic gradient followed by a resonant light pulse that removes atoms in the remaining layers. Next, we prepare a 50/50 mixture of $\left|9/2, -9/2\ket$ and $\left|9/2, -7/2\ket$ to allow thermalization, and transfer the atoms to a vertical ($z$ direction) $1064\,\rm nm$ beam, forming a lattice along $z$ with a spacing of $532\,\rm nm$ (Fig.~1(a)). After evaporating by lowering the power of the $z$-lattice, the $z$-depth is increased to $180\, \mu\rm K$. We simultaneously ramp up two additional $1064\,\rm nm$ beams (Fig.~1(a)) reflected off the substrate at $10.8^{\circ}$ and retro-reflected. These form a lattice in the horizontal plane with a spacing of $541\,\rm nm$.

\begin{figure}
\centering
\includegraphics[scale=1.0]{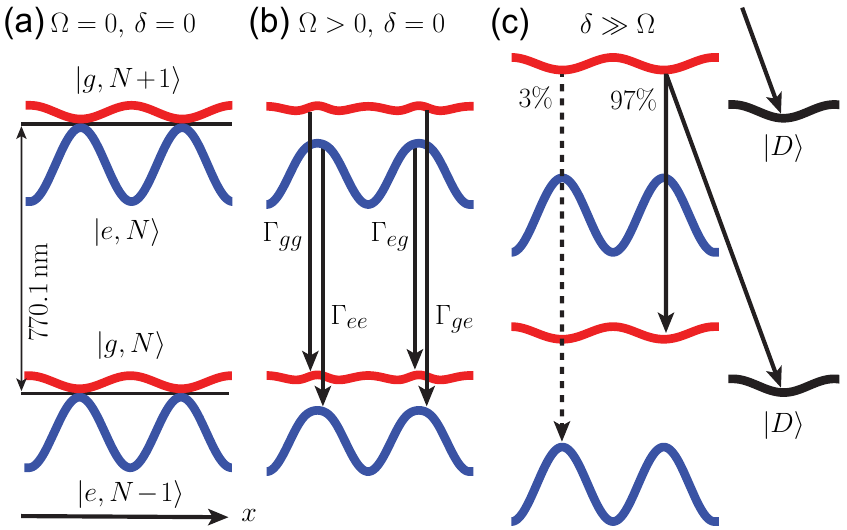}
\caption{(a) Electronic ground ($\left | g, N\ket$) and excited ($\left| e,N\ket$) states dressed by $N$ optical pumping photons. (b) On resonance, all dressed states are equally populated and experience an anti-trapping potential. The four decay channels between the dressed states are equal. (c) The dressed states with $\Omega/\delta= 0.175$, shown with the dark state $\left| D \ket$. Only a small fraction $\Omega^4/\delta^4$ of the steady-state population resides in anti-trapping states. For the atoms in the trapping state, the branching ratios between trapping transitions (solid arrows) and anti-trapping transitions (dashed arrow) are shown. }
\label{fig2}
\end{figure}

\begin{figure*}
\centering
\includegraphics[scale=1.0]{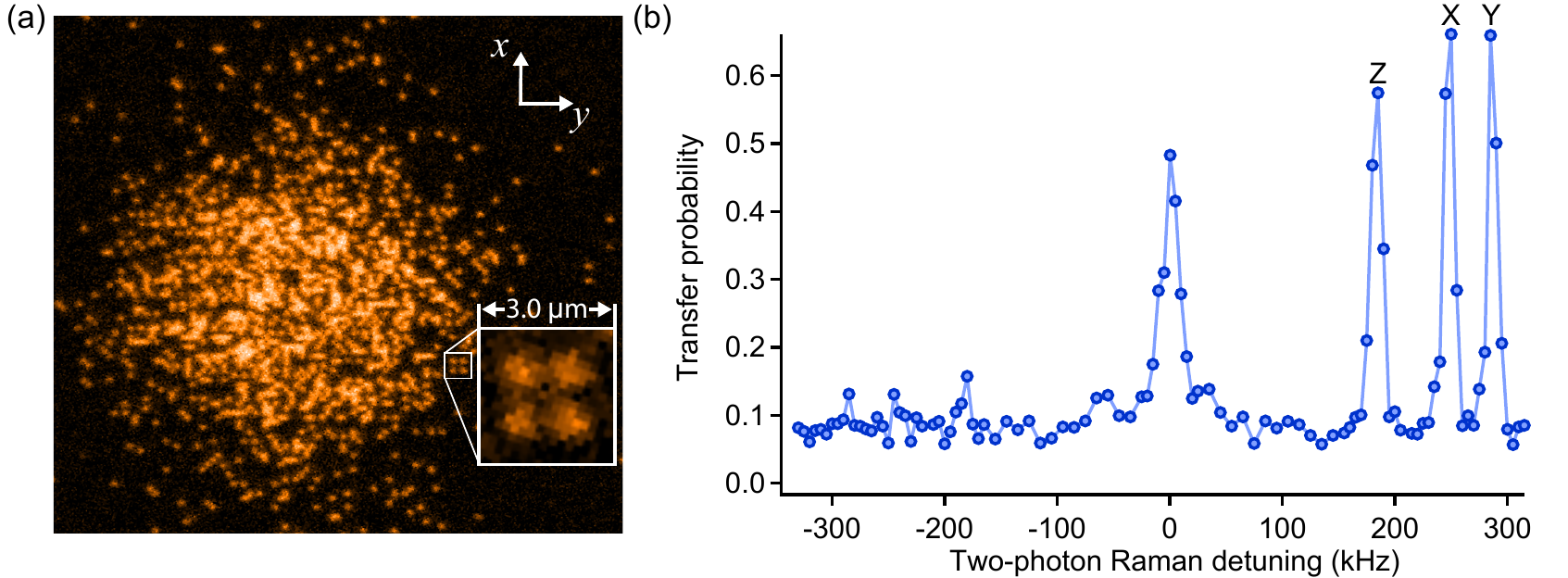}
\caption{(a) Site-resolved imaging of fermonic atoms on a densely-filled $541\,\rm nm$-period optical lattice, with an exposure time of two seconds; one can clearly discern the lattice structure and individual atoms. (b) Raman spectrum after cooling. The lattice depths are chosen such that the vibrational sidebands are well-resolved. The heating sidebands for the three lattice axes are labeled Z, X and Y. We observe a large sideband asymmetry, from which we extract a 3D ground state occupation of $72(3)\%$.}
\label{fig3}
\end{figure*}

During imaging, the atoms are trapped in a deep lattice, where the potential at each lattice site can be approximated by a harmonic well with vibrational frequency $\omega$. At our imaging depth, the vibrational frequencies for the three axes are $(\omega_x, \omega_y, \omega_z) =2\pi \times (280,300,260)\,\rm kHz$, corresponding to lattice depths of $210\,\mu\rm K$, $240\,\mu\rm K$ and $180\,\mu\rm K$ respectively. The Rabi coupling for transitions that change the vibrational number by one is proportional to the Lamb-Dicke parameter, $\eta = \Delta k \, a$, where $a = \sqrt{\frac{\hbar}{2m\omega}}$ is the harmonic oscillator length and $\hbar \Delta k$ is the momentum transfer due to the Raman beams. Along our lattice directions, $\Delta k_x=8.0\,\mu \rm m^{-1}$, $\Delta k_y=8.0\,\mu\rm m^{-1}$ and $\Delta k_z = 3.1\,\mu \rm m^{-1}$, yielding Lamb-Dicke parameters of 0.17 for $x$ and $y$, and 0.068 for $z$. The polarizations of both Raman beams are linear and parallel to the substrate (Fig.~1(b)), in order to avoid differential effective magnetic fields between $\left|a\ket$ and $\left|b\ket$ that would arise for circularly polarized light. The Raman beam along the $y$-axis contains a single frequency, whereas the Raman beam along the $x$-axis contains three frequencies, allowing us to address the cooling sidebands of the three directions simultaneously. The resulting two-photon detunings from the bare $\left |a\ket \rightarrow \left |b\ket$ transition are $400\,\rm kHz$, $450\,\rm kHz$ and $360\,\rm kHz$ for cooling along $x$, $y$ and $z$ respectively. These frequencies compensate for differential Stark shifts that arise in the presence of optical pumping light. The $x$-Raman beam intensities of the three frequency components are $0.79\, \rm W/cm^2$, $0.47\, \rm W/cm^2$ and $0.49\,\rm W/cm^2$ respectively; the intensity of the $y$-Raman beam is $2.0\,\rm W/cm^2$.

In addition to these Raman beams, optical pumping light is present to complete the cooling cycle. During optical pumping, atoms enter electronically excited states, and preferentially decay into the desired state. Typically, the excited states experience an anti-trapping potential when the ground state experiences a trapping potential. For our $1064\,\rm nm$ lattice, the anti-trapping potential for the atoms in the $\rm 4\, P_{1/2}$ states is $5.4$ times stronger than the trapping potential for atoms in the $\rm 4\,S_{1/2}$ states, due to the $\rm 4\, P_{1/2} \rightarrow\rm3\,D_{3/2}$ transition at $1169\,\rm nm$. This strong anti-trapping would lead to heating and diffusion of atoms through the lattice during imaging.

A solution to this problem is to detune the optical pumping light away from resonance. This reduces the population in the anti-trapping states and favors transitions into trapping states. One can understand this by considering the dressed states of a driven two-level system. At zero intensity, the dressed states, $\left|g,N\ket$ and $\left|e,N\ket$, correspond to the bare trapping and anti-trapping states with $N$ photons respectively (Fig.~2(a)). In the presence of resonant pumping light, neither of the dressed states is trapping (Fig.~2(b)), which leads to heating.

\begin{figure*}
\centering
\includegraphics[scale=1.0]{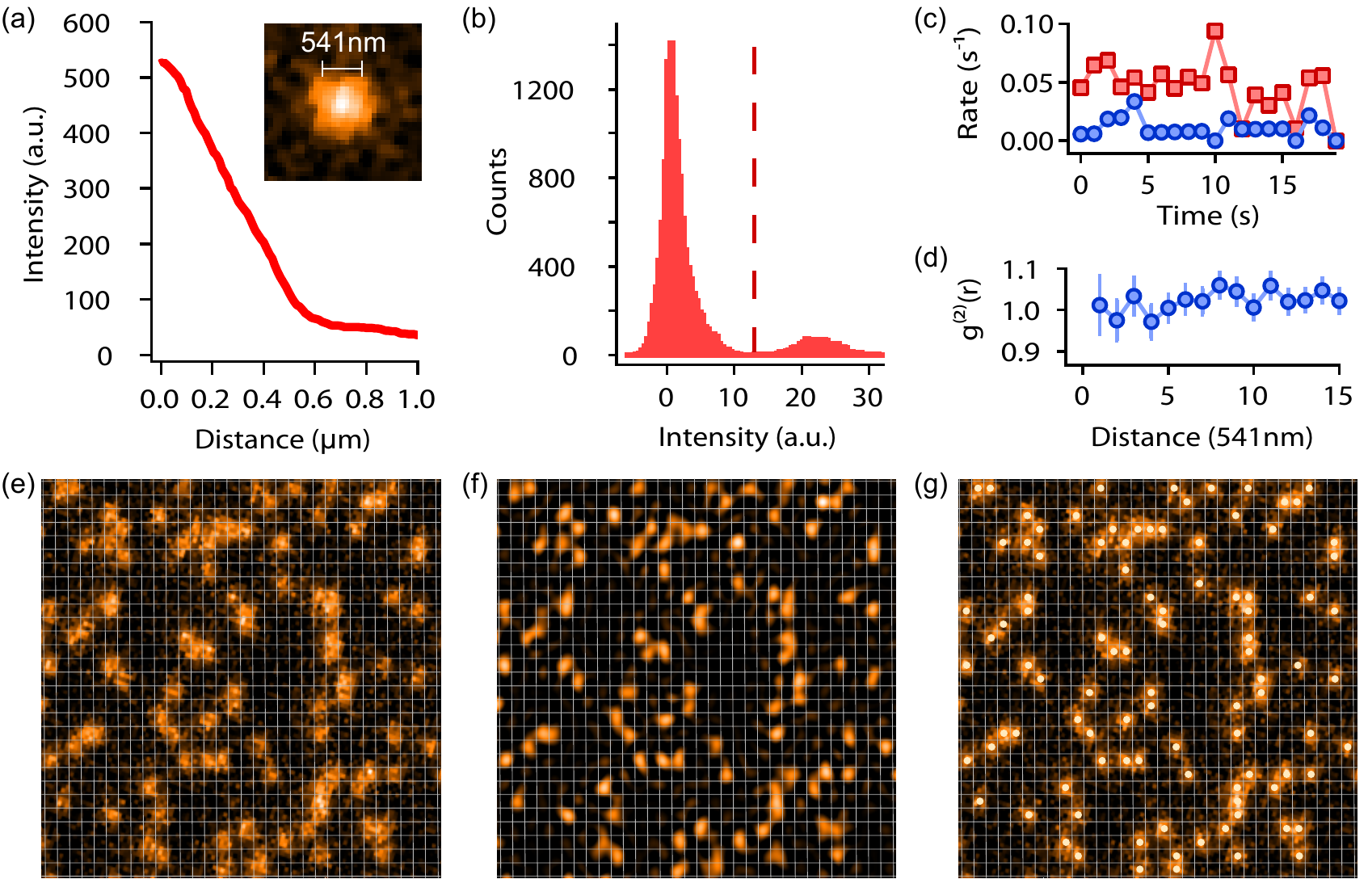}
\caption{(a) Radially averaged PSF extracted from isolated atoms; the FWHM is $640\,\rm nm$. (b) Intensity histogram after binning the deconvoluted images by lattice site. The threshold for reconstruction is shown by the dashed line. (c) Loss and hopping rates, shown in red squares and blue circles respectively, as extracted from 20 consecutive one-second exposures. (d) Correlation measurement $g^{(2)}(r)$ of a thermal cloud, showing the absence of distance-dependent loss. (e) Image of sparsely filled lattice with grid lines showing lattice spacing and orientation. The exposure time is one second. (f,g) The same image after deconvolution and with the filled sites identified.}
\label{fig4}
\end{figure*}

However, at large detunings $\delta \gg \Omega $, where $\Omega$ is the Rabi frequency, one dressed state becomes trapping. Furthermore, spontaneous decay among the dressed states favors population in the trapping states. Specifically, the decay rates $\Gamma_{eg}$, $\Gamma_{ee}$, $\Gamma_{gg}$ and $\Gamma_{ge}$, defined in Fig.~2(b), are proportional to $1$, $s$, $s$ and $s^2$ respectively, where $s =\Omega^2/\delta^2$. The ratio of the anti-trapped population to the trapped population is suppressed, because in steady state, it is given by $\Gamma_{ge}/\Gamma_{eg}$ which scales as $s^2$. Another benefit of large detunings involves the state $\left|D\ket$ into which atoms are optically pumped (Fig.~2(c)). This state is dark to the optical pumping light, and hence has no excited state admixture; consequently, it experiences a trapping potential. Atoms in trapping states $\left|g,N\ket$ decay preferentially into $\left|D\ket$, since the ratio of anti-trapping transitions to dark state transitions scales as $\Gamma_{ge}/\Gamma_{gg} = s$, which is small at large $\delta$.

In light of these considerations, we detune the hyperfine-changing ($F$) pumping light $-80\,\rm MHz$ from the Stark-shifted $F=7/2 \rightarrow F'=9/2$ transition and the Zeeman-level ($m_F$) pumping light $-80\,\rm MHz$ from the Stark-shifted $F=9/2 \rightarrow F'=7/2$ transition. The optical pumping beam co-propagates with the $x$-Raman beam, and has its polarization optimized for minimal $\sigma_+$ admixture. The $F$ and $m_F$ components have intensities of $5.8\,\rm mW/cm^2$ and $1.6\,\rm mW/cm^2$ respectively; the Lamb-Dicke parameters for optical pumping are $ 0.17$ for $x$ and $y$, and 0.18 for $z$.

This Raman cooling scheme allows us to collect fluorescence while keeping the atoms confined to their lattice sites (Fig.~3(a)). Furthermore, we find that atoms are cooled predominantly into their motional ground state. Indeed, Raman spectroscopy reveals a 3D ground state population of $72(3)\%$ after cooling (Fig.~3(b)). Note that parameters are optimized for imaging fidelity rather than for a large ground state population. We measure a fluorescence rate of $\sim 5000\,\rm photons/atom \cdot s$ with a lifetime of $\sim30$ seconds. With a photon collection and detection efficiency of $20\%$, about $1000$ photons per atom can be collected in an exposure time of one second, which is sufficient to detect single atoms with high fidelity. Using the number of scattered photons and the Lamb-Dicke parameters for spontaneous emission, we estimate the cooling rate to be $\sim 4\, \rm \mu \rm K/ms$.

To verify that we can resolve individual lattice sites, we measure the point spread function (PSF) of our imaging system using isolated atoms from sparsely populated images. The measured PSF has a full-width-half-max (FWHM) of $640\,\rm nm$ (Fig.~4(a)). Images are deconvolved with the PSF to achieve sub-lattice-site resolution. From such images, we also extract the lattice axes and spacings necessary to reconstruct the atomic distribution. Binning the intensity of the deconvolved image by lattice site reveals a clear bimodal distribution (Fig.~4(b)), which is used to determine whether a site is filled (Fig.~4(e,f,g)). This bimodality gives a reconstruction error of $<\%1$.

An important aspect of quantum gas microscopy is the fidelity of the imaging process, which can be characterized by hopping and loss rates. To this end, we take a series of images of the same atomic cloud and observe changes in the site occupations between images. Sites that are empty but become occupied in a subsequent image are counted as hopping events; sites that become empty are counted as loss events. The Raman cooling parameters are optimized for low hopping and loss rates while maintaining a fixed level of fluorescence. For optimized parameters, we achieve loss rates of $<4.8\pm0.2\%$ and hopping rates of $<1.2\pm0.2\%$ for one-second exposures of clouds with fillings between 0.10 and 0.20 (Fig.~4(c)). These rates, which include reconstruction errors, give an detection fidelity of $>95\%$ for sparse clouds. At higher fillings, hopping events lead to loss of additional atoms in doubly-occupied sites due to light-assisted collisions. However, even for unity filling, we estimate the imaging fidelity to still be $>94\%$, because of the low hopping rate.

To ensure that the imaging does not cause additional losses for neighboring atoms, one can measure the 2-point correlation function $g^{(2)}(r) =\left \bra n(x) n(x+r)\ket/\left \bra n\ket^2$ of thermal clouds, since distance-dependent loss will produce anti-correlations at short distances. For a dilute thermal cloud with a filling of 0.19, one $29\times32$ site image gives $g^{(2)}(r)=1.00(7)$ for distances from $r=1$ to 10 lattice spacings, indicating that the imaging does not cause significant distance-dependent loss (Fig.~4(d)).

In conclusion, we have realized high-fidelity site-resolved imaging of $^{40}$K fermionic atoms in a Hubbard-type optical lattice by combining 3D Raman sideband cooling with high resolution photon collection. In contrast to existing boson microscopes, the technique leaves atoms predominantly in the absolute 3D ground state of a given lattice site. This opens up new ways to assemble low-entropy Fermi-Hubbard systems atom by atom~\cite{weiss2004, murmann2015,kaufman2014HOM}. Combining site-resolved imaging with on-site manipulation would allow one to deterministically create localized excitations and follow their time evolution~\cite{Cheneau2012Lightcone}. Finally, the presence of $^{23}$Na in our system invites the realization of a quantum gas microscope for ultracold fermionic NaK molecules~\cite{wu2012NaK}, which have been proposed as a new resource for quantum information processing and quantum simulation of lattice models with long-range dipolar interactions.

We would like to thank Katherine Lawrence for experimental assistance and critical readings of the manuscript. This work was supported by the NSF, AFOSR-PECASE, AFOSR-MURI on Exotic Phases of Matter, ARO-MURI on Atomtronics, ONR, a grant from the Army Research Office with funding from the DARPA OLE program, and the David and Lucile Packard Foundation.

\bibliographystyle{apsrev4-1}

\begin{thebibliography}{31}%
\makeatletter
\providecommand \@ifxundefined [1]{%
 \@ifx{#1\undefined}
}%
\providecommand \@ifnum [1]{%
 \ifnum #1\expandafter \@firstoftwo
 \else \expandafter \@secondoftwo
 \fi
}%
\providecommand \@ifx [1]{%
 \ifx #1\expandafter \@firstoftwo
 \else \expandafter \@secondoftwo
 \fi
}%
\providecommand \natexlab [1]{#1}%
\providecommand \enquote  [1]{``#1''}%
\providecommand \bibnamefont  [1]{#1}%
\providecommand \bibfnamefont [1]{#1}%
\providecommand \citenamefont [1]{#1}%
\providecommand \href@noop [0]{\@secondoftwo}%
\providecommand \href [0]{\begingroup \@sanitize@url \@href}%
\providecommand \@href[1]{\@@startlink{#1}\@@href}%
\providecommand \@@href[1]{\endgroup#1\@@endlink}%
\providecommand \@sanitize@url [0]{\catcode `\\12\catcode `\$12\catcode
  `\&12\catcode `\#12\catcode `\^12\catcode `\_12\catcode `\%12\relax}%
\providecommand \@@startlink[1]{}%
\providecommand \@@endlink[0]{}%
\providecommand \url  [0]{\begingroup\@sanitize@url \@url }%
\providecommand \@url [1]{\endgroup\@href {#1}{\urlprefix }}%
\providecommand \urlprefix  [0]{URL }%
\providecommand \Eprint [0]{\href }%
\providecommand \doibase [0]{http://dx.doi.org/}%
\providecommand \selectlanguage [0]{\@gobble}%
\providecommand \bibinfo  [0]{\@secondoftwo}%
\providecommand \bibfield  [0]{\@secondoftwo}%
\providecommand \translation [1]{[#1]}%
\providecommand \BibitemOpen [0]{}%
\providecommand \bibitemStop [0]{}%
\providecommand \bibitemNoStop [0]{.\EOS\space}%
\providecommand \EOS [0]{\spacefactor3000\relax}%
\providecommand \BibitemShut  [1]{\csname bibitem#1\endcsname}%
\let\auto@bib@innerbib\@empty
\bibitem [{\citenamefont {Inguscio}\ \emph {et~al.}(2008)\citenamefont
  {Inguscio}, \citenamefont {Ketterle},\ and\ \citenamefont
  {Salomon}}]{ingu08varenna}%
  \BibitemOpen
  \bibinfo {editor} {\bibfnamefont {M.}~\bibnamefont {Inguscio}}, \bibinfo
  {editor} {\bibfnamefont {W.}~\bibnamefont {Ketterle}}, \ and\ \bibinfo
  {editor} {\bibfnamefont {C.}~\bibnamefont {Salomon}},\ eds.,\ \href@noop {}
  {\emph {\bibinfo {title} {Ultracold Fermi Gases}}},\ Proceedings of the
  International School of Physics "Enrico Fermi", Course CLXIV, Varenna, 20 -
  30 June 2006\ (\bibinfo  {publisher} {IOS Press, Amsterdam},\ \bibinfo {year}
  {2008})\BibitemShut {NoStop}%
\bibitem [{\citenamefont {Bloch}\ \emph {et~al.}(2008)\citenamefont {Bloch},
  \citenamefont {Dalibard},\ and\ \citenamefont {Zwerger}}]{bloc08review}%
  \BibitemOpen
  \bibfield  {author} {\bibinfo {author} {\bibfnamefont {I.}~\bibnamefont
  {Bloch}}, \bibinfo {author} {\bibfnamefont {J.}~\bibnamefont {Dalibard}}, \
  and\ \bibinfo {author} {\bibfnamefont {W.}~\bibnamefont {Zwerger}},\
  }\href@noop {} {\bibfield  {journal} {\bibinfo  {journal} {Rev. Mod. Phys.}\
  }\textbf {\bibinfo {volume} {80}},\ \bibinfo {pages} {885} (\bibinfo {year}
  {2008})}\BibitemShut {NoStop}%
\bibitem [{\citenamefont {Zwerger}(2011)}]{Zwerger2011BECBCS}%
  \BibitemOpen
  \bibinfo {editor} {\bibfnamefont {W.}~\bibnamefont {Zwerger}},\ ed.,\
  \href@noop {} {\emph {\bibinfo {title} {The BCS-BEC crossover and the unitary
  {Fermi} gas}}},\ Vol.\ \bibinfo {volume} {836}\ (\bibinfo  {publisher}
  {Springer},\ \bibinfo {year} {2011})\BibitemShut {NoStop}%
\bibitem [{\citenamefont {Zwierlein}(2014)}]{Zwierlein2014NovelSuperfluids}%
  \BibitemOpen
  \bibfield  {author} {\bibinfo {author} {\bibfnamefont {M.~W.}\ \bibnamefont
  {Zwierlein}},\ }in\ \href@noop {} {\emph {\bibinfo {booktitle} {Novel
  Superfluids, Vol. 2}}},\ \bibinfo {editor} {edited by\ \bibinfo {editor}
  {\bibfnamefont {K.-H.}\ \bibnamefont {Bennemann}}\ and\ \bibinfo {editor}
  {\bibfnamefont {J.~B.}\ \bibnamefont {Ketterson}}}\ (\bibinfo  {publisher}
  {Oxford University Press},\ \bibinfo {address} {Oxford},\ \bibinfo {year}
  {2014})\BibitemShut {NoStop}%
\bibitem [{\citenamefont {Esslinger}(2010)}]{Esslinger2010FermiHubbard}%
  \BibitemOpen
  \bibfield  {author} {\bibinfo {author} {\bibfnamefont {T.}~\bibnamefont
  {Esslinger}},\ }\href@noop {} {\bibfield  {journal} {\bibinfo  {journal}
  {Annual Review of Condensed Matter Physics}\ }\textbf {\bibinfo {volume}
  {1}},\ \bibinfo {pages} {129} (\bibinfo {year} {2010})}\BibitemShut {NoStop}%
\bibitem [{\citenamefont {Chin}\ \emph {et~al.}(2006)\citenamefont {Chin},
  \citenamefont {Miller}, \citenamefont {Liu}, \citenamefont {Stan},
  \citenamefont {Setiawan}, \citenamefont {Sanner}, \citenamefont {Xu},\ and\
  \citenamefont {Ketterle}}]{chin06}%
  \BibitemOpen
  \bibfield  {author} {\bibinfo {author} {\bibfnamefont {J.}~\bibnamefont
  {Chin}}, \bibinfo {author} {\bibfnamefont {D.}~\bibnamefont {Miller}},
  \bibinfo {author} {\bibfnamefont {Y.}~\bibnamefont {Liu}}, \bibinfo {author}
  {\bibfnamefont {C.}~\bibnamefont {Stan}}, \bibinfo {author} {\bibfnamefont
  {W.}~\bibnamefont {Setiawan}}, \bibinfo {author} {\bibfnamefont
  {C.}~\bibnamefont {Sanner}}, \bibinfo {author} {\bibfnamefont
  {K.}~\bibnamefont {Xu}}, \ and\ \bibinfo {author} {\bibfnamefont
  {W.}~\bibnamefont {Ketterle}},\ }\href@noop {} {\bibfield  {journal}
  {\bibinfo  {journal} {Nature}\ }\textbf {\bibinfo {volume} {443}},\ \bibinfo
  {pages} {961} (\bibinfo {year} {2006})}\BibitemShut {NoStop}%
\bibitem [{\citenamefont {J\"ordens}\ \emph {et~al.}(2008)\citenamefont
  {J\"ordens}, \citenamefont {Strohmaier}, \citenamefont {G\"unter},
  \citenamefont {Moritz},\ and\ \citenamefont {Esslinger}}]{jord08}%
  \BibitemOpen
  \bibfield  {author} {\bibinfo {author} {\bibfnamefont {R.}~\bibnamefont
  {J\"ordens}}, \bibinfo {author} {\bibfnamefont {N.}~\bibnamefont
  {Strohmaier}}, \bibinfo {author} {\bibfnamefont {K.}~\bibnamefont
  {G\"unter}}, \bibinfo {author} {\bibfnamefont {H.}~\bibnamefont {Moritz}}, \
  and\ \bibinfo {author} {\bibfnamefont {T.}~\bibnamefont {Esslinger}},\
  }\href@noop {} {\bibfield  {journal} {\bibinfo  {journal} {Nature}\ }\textbf
  {\bibinfo {volume} {455}},\ \bibinfo {pages} {204} (\bibinfo {year}
  {2008})}\BibitemShut {NoStop}%
\bibitem [{\citenamefont {Schneider}\ \emph {et~al.}(2008)\citenamefont
  {Schneider}, \citenamefont {Hackerm\"uller}, \citenamefont {Will},
  \citenamefont {Best}, \citenamefont {Bloch}, \citenamefont {Costi},
  \citenamefont {Helmes}, \citenamefont {Rasch},\ and\ \citenamefont
  {Rosch}}]{schn08}%
  \BibitemOpen
  \bibfield  {author} {\bibinfo {author} {\bibfnamefont {U.}~\bibnamefont
  {Schneider}}, \bibinfo {author} {\bibfnamefont {L.}~\bibnamefont
  {Hackerm\"uller}}, \bibinfo {author} {\bibfnamefont {S.}~\bibnamefont
  {Will}}, \bibinfo {author} {\bibfnamefont {T.}~\bibnamefont {Best}}, \bibinfo
  {author} {\bibfnamefont {I.}~\bibnamefont {Bloch}}, \bibinfo {author}
  {\bibfnamefont {T.~A.}\ \bibnamefont {Costi}}, \bibinfo {author}
  {\bibfnamefont {R.~W.}\ \bibnamefont {Helmes}}, \bibinfo {author}
  {\bibfnamefont {D.}~\bibnamefont {Rasch}}, \ and\ \bibinfo {author}
  {\bibfnamefont {A.}~\bibnamefont {Rosch}},\ }\href@noop {} {\bibfield
  {journal} {\bibinfo  {journal} {Science}\ }\textbf {\bibinfo {volume}
  {322}},\ \bibinfo {pages} {1520} (\bibinfo {year} {2008})}\BibitemShut
  {NoStop}%
\bibitem [{\citenamefont {Greif}\ \emph {et~al.}(2013)\citenamefont {Greif},
  \citenamefont {Uehlinger}, \citenamefont {Jotzu}, \citenamefont {Tarruell},\
  and\ \citenamefont {Esslinger}}]{Greif2013Magnetism}%
  \BibitemOpen
  \bibfield  {author} {\bibinfo {author} {\bibfnamefont {D.}~\bibnamefont
  {Greif}}, \bibinfo {author} {\bibfnamefont {T.}~\bibnamefont {Uehlinger}},
  \bibinfo {author} {\bibfnamefont {G.}~\bibnamefont {Jotzu}}, \bibinfo
  {author} {\bibfnamefont {L.}~\bibnamefont {Tarruell}}, \ and\ \bibinfo
  {author} {\bibfnamefont {T.}~\bibnamefont {Esslinger}},\ }\href@noop {}
  {\bibfield  {journal} {\bibinfo  {journal} {Science}\ }\textbf {\bibinfo
  {volume} {340}},\ \bibinfo {pages} {1307} (\bibinfo {year}
  {2013})}\BibitemShut {NoStop}%
\bibitem [{\citenamefont {Imri\v{s}ka}\ \emph {et~al.}(2014)\citenamefont
  {Imri\v{s}ka}, \citenamefont {Iazzi}, \citenamefont {Wang}, \citenamefont
  {Gull}, \citenamefont {Greif}, \citenamefont {Uehlinger}, \citenamefont
  {Jotzu}, \citenamefont {Tarruell}, \citenamefont {Esslinger},\ and\
  \citenamefont {Troyer}}]{Imriska2014Hubbard}%
  \BibitemOpen
  \bibfield  {author} {\bibinfo {author} {\bibfnamefont {J.}~\bibnamefont
  {Imri\v{s}ka}}, \bibinfo {author} {\bibfnamefont {M.}~\bibnamefont {Iazzi}},
  \bibinfo {author} {\bibfnamefont {L.}~\bibnamefont {Wang}}, \bibinfo {author}
  {\bibfnamefont {E.}~\bibnamefont {Gull}}, \bibinfo {author} {\bibfnamefont
  {D.}~\bibnamefont {Greif}}, \bibinfo {author} {\bibfnamefont
  {T.}~\bibnamefont {Uehlinger}}, \bibinfo {author} {\bibfnamefont
  {G.}~\bibnamefont {Jotzu}}, \bibinfo {author} {\bibfnamefont
  {L.}~\bibnamefont {Tarruell}}, \bibinfo {author} {\bibfnamefont
  {T.}~\bibnamefont {Esslinger}}, \ and\ \bibinfo {author} {\bibfnamefont
  {M.}~\bibnamefont {Troyer}},\ }\href@noop {} {\bibfield  {journal} {\bibinfo
  {journal} {Phys. Rev. Lett.}\ }\textbf {\bibinfo {volume} {112}},\ \bibinfo
  {pages} {115301} (\bibinfo {year} {2014})}\BibitemShut {NoStop}%
\bibitem [{\citenamefont {Hart}\ \emph {et~al.}(2014)\citenamefont {Hart},
  \citenamefont {Duarte}, \citenamefont {Yang}, \citenamefont {Liu},
  \citenamefont {Paiva}, \citenamefont {Khatami}, \citenamefont {Scalettar},
  \citenamefont {Trivedi}, \citenamefont {Huse},\ and\ \citenamefont
  {Hulet}}]{Hart2014FermiHubbard}%
  \BibitemOpen
  \bibfield  {author} {\bibinfo {author} {\bibfnamefont {R.~A.}\ \bibnamefont
  {Hart}}, \bibinfo {author} {\bibfnamefont {P.~M.}\ \bibnamefont {Duarte}},
  \bibinfo {author} {\bibfnamefont {T.~L.}\ \bibnamefont {Yang}}, \bibinfo
  {author} {\bibfnamefont {X.~X.}\ \bibnamefont {Liu}}, \bibinfo {author}
  {\bibfnamefont {T.}~\bibnamefont {Paiva}}, \bibinfo {author} {\bibfnamefont
  {E.}~\bibnamefont {Khatami}}, \bibinfo {author} {\bibfnamefont
  {R.}~\bibnamefont {Scalettar}}, \bibinfo {author} {\bibfnamefont
  {N.}~\bibnamefont {Trivedi}}, \bibinfo {author} {\bibfnamefont {D.~A.}\
  \bibnamefont {Huse}}, \ and\ \bibinfo {author} {\bibfnamefont {R.~G.}\
  \bibnamefont {Hulet}},\ }\href@noop {} {\bibfield  {journal} {\bibinfo
  {journal} {Preprint arXiv:1407.5932}\ } (\bibinfo {year} {2014})}\BibitemShut
  {NoStop}%
\bibitem [{\citenamefont {Bakr}\ \emph {et~al.}(2009)\citenamefont {Bakr},
  \citenamefont {Gillen}, \citenamefont {Peng}, \citenamefont {F\"olling},\
  and\ \citenamefont {Greiner}}]{bakr2009microscope}%
  \BibitemOpen
  \bibfield  {author} {\bibinfo {author} {\bibfnamefont {W.~S.}\ \bibnamefont
  {Bakr}}, \bibinfo {author} {\bibfnamefont {J.~I.}\ \bibnamefont {Gillen}},
  \bibinfo {author} {\bibfnamefont {A.}~\bibnamefont {Peng}}, \bibinfo {author}
  {\bibfnamefont {S.}~\bibnamefont {F\"olling}}, \ and\ \bibinfo {author}
  {\bibfnamefont {M.}~\bibnamefont {Greiner}},\ }\href@noop {} {\bibfield
  {journal} {\bibinfo  {journal} {Nature}\ }\textbf {\bibinfo {volume} {462}},\
  \bibinfo {pages} {74} (\bibinfo {year} {2009})}\BibitemShut {NoStop}%
\bibitem [{\citenamefont {Sherson}\ \emph {et~al.}(2010)\citenamefont
  {Sherson}, \citenamefont {Weitenberg}, \citenamefont {Endres}, \citenamefont
  {Cheneau}, \citenamefont {Bloch},\ and\ \citenamefont
  {Kuhr}}]{sherson2010microscope}%
  \BibitemOpen
  \bibfield  {author} {\bibinfo {author} {\bibfnamefont {J.~F.}\ \bibnamefont
  {Sherson}}, \bibinfo {author} {\bibfnamefont {C.}~\bibnamefont {Weitenberg}},
  \bibinfo {author} {\bibfnamefont {M.}~\bibnamefont {Endres}}, \bibinfo
  {author} {\bibfnamefont {M.}~\bibnamefont {Cheneau}}, \bibinfo {author}
  {\bibfnamefont {I.}~\bibnamefont {Bloch}}, \ and\ \bibinfo {author}
  {\bibfnamefont {S.}~\bibnamefont {Kuhr}},\ }\href@noop {} {\bibfield
  {journal} {\bibinfo  {journal} {Nature}\ }\textbf {\bibinfo {volume} {467}},\
  \bibinfo {pages} {68} (\bibinfo {year} {2010})}\BibitemShut {NoStop}%
\bibitem [{\citenamefont {{Miranda}}\ \emph {et~al.}(2014)\citenamefont
  {{Miranda}}, \citenamefont {{Inoue}}, \citenamefont {{Okuyama}},
  \citenamefont {{Nakamoto}},\ and\ \citenamefont {{Kozuma}}}]{miranda2014}%
  \BibitemOpen
  \bibfield  {author} {\bibinfo {author} {\bibfnamefont {M.}~\bibnamefont
  {{Miranda}}}, \bibinfo {author} {\bibfnamefont {R.}~\bibnamefont {{Inoue}}},
  \bibinfo {author} {\bibfnamefont {Y.}~\bibnamefont {{Okuyama}}}, \bibinfo
  {author} {\bibfnamefont {A.}~\bibnamefont {{Nakamoto}}}, \ and\ \bibinfo
  {author} {\bibfnamefont {M.}~\bibnamefont {{Kozuma}}},\ }\href@noop {}
  {\bibfield  {journal} {\bibinfo  {journal} {arXiv:1410.5189}\ } (\bibinfo
  {year} {2014})}\BibitemShut {NoStop}%
\bibitem [{\citenamefont {Bakr}\ \emph {et~al.}(2010)\citenamefont {Bakr},
  \citenamefont {Peng}, \citenamefont {Tai}, \citenamefont {Ma}, \citenamefont
  {Simon}, \citenamefont {Gillen}, \citenamefont {F\"olling}, \citenamefont
  {Pollet},\ and\ \citenamefont {Greiner}}]{bakr2010MottInsulator}%
  \BibitemOpen
  \bibfield  {author} {\bibinfo {author} {\bibfnamefont {W.~S.}\ \bibnamefont
  {Bakr}}, \bibinfo {author} {\bibfnamefont {A.}~\bibnamefont {Peng}}, \bibinfo
  {author} {\bibfnamefont {M.~E.}\ \bibnamefont {Tai}}, \bibinfo {author}
  {\bibfnamefont {R.}~\bibnamefont {Ma}}, \bibinfo {author} {\bibfnamefont
  {J.}~\bibnamefont {Simon}}, \bibinfo {author} {\bibfnamefont {J.~I.}\
  \bibnamefont {Gillen}}, \bibinfo {author} {\bibfnamefont {S.}~\bibnamefont
  {F\"olling}}, \bibinfo {author} {\bibfnamefont {L.}~\bibnamefont {Pollet}}, \
  and\ \bibinfo {author} {\bibfnamefont {M.}~\bibnamefont {Greiner}},\
  }\href@noop {} {\bibfield  {journal} {\bibinfo  {journal} {Science}\ }\textbf
  {\bibinfo {volume} {329}},\ \bibinfo {pages} {547} (\bibinfo {year}
  {2010})}\BibitemShut {NoStop}%
\bibitem [{\citenamefont {Endres}\ \emph {et~al.}(2011)\citenamefont {Endres},
  \citenamefont {Cheneau}, \citenamefont {Fukuhara}, \citenamefont
  {Weitenberg}, \citenamefont {Schau\ss}, \citenamefont {Gross}, \citenamefont
  {Mazza}, \citenamefont {Banuls}, \citenamefont {Pollet}, \citenamefont
  {Bloch},\ and\ \citenamefont {Kuhr}}]{Endres2011stringorder}%
  \BibitemOpen
  \bibfield  {author} {\bibinfo {author} {\bibfnamefont {M.}~\bibnamefont
  {Endres}}, \bibinfo {author} {\bibfnamefont {M.}~\bibnamefont {Cheneau}},
  \bibinfo {author} {\bibfnamefont {T.}~\bibnamefont {Fukuhara}}, \bibinfo
  {author} {\bibfnamefont {C.}~\bibnamefont {Weitenberg}}, \bibinfo {author}
  {\bibfnamefont {P.}~\bibnamefont {Schau\ss}}, \bibinfo {author}
  {\bibfnamefont {C.}~\bibnamefont {Gross}}, \bibinfo {author} {\bibfnamefont
  {L.}~\bibnamefont {Mazza}}, \bibinfo {author} {\bibfnamefont {M.~C.}\
  \bibnamefont {Banuls}}, \bibinfo {author} {\bibfnamefont {L.}~\bibnamefont
  {Pollet}}, \bibinfo {author} {\bibfnamefont {I.}~\bibnamefont {Bloch}}, \
  and\ \bibinfo {author} {\bibfnamefont {S.}~\bibnamefont {Kuhr}},\ }\href@noop
  {} {\bibfield  {journal} {\bibinfo  {journal} {Science}\ }\textbf {\bibinfo
  {volume} {334}},\ \bibinfo {pages} {200} (\bibinfo {year}
  {2011})}\BibitemShut {NoStop}%
\bibitem [{\citenamefont {Cheneau}\ \emph {et~al.}(2012)\citenamefont
  {Cheneau}, \citenamefont {Barmettler}, \citenamefont {Poletti}, \citenamefont
  {Endres}, \citenamefont {Schau\ss}, \citenamefont {Fukuhara}, \citenamefont
  {Gross}, \citenamefont {Bloch}, \citenamefont {Kollath},\ and\ \citenamefont
  {Kuhr}}]{Cheneau2012Lightcone}%
  \BibitemOpen
  \bibfield  {author} {\bibinfo {author} {\bibfnamefont {M.}~\bibnamefont
  {Cheneau}}, \bibinfo {author} {\bibfnamefont {P.}~\bibnamefont {Barmettler}},
  \bibinfo {author} {\bibfnamefont {D.}~\bibnamefont {Poletti}}, \bibinfo
  {author} {\bibfnamefont {M.}~\bibnamefont {Endres}}, \bibinfo {author}
  {\bibfnamefont {P.}~\bibnamefont {Schau\ss}}, \bibinfo {author}
  {\bibfnamefont {T.}~\bibnamefont {Fukuhara}}, \bibinfo {author}
  {\bibfnamefont {C.}~\bibnamefont {Gross}}, \bibinfo {author} {\bibfnamefont
  {I.}~\bibnamefont {Bloch}}, \bibinfo {author} {\bibfnamefont
  {C.}~\bibnamefont {Kollath}}, \ and\ \bibinfo {author} {\bibfnamefont
  {S.}~\bibnamefont {Kuhr}},\ }\href@noop {} {\bibfield  {journal} {\bibinfo
  {journal} {Nature}\ }\textbf {\bibinfo {volume} {481}},\ \bibinfo {pages}
  {484} (\bibinfo {year} {2012})}\BibitemShut {NoStop}%
\bibitem [{\citenamefont {Monroe}\ \emph {et~al.}(1995)\citenamefont {Monroe},
  \citenamefont {Meekhof}, \citenamefont {King}, \citenamefont {Jefferts},
  \citenamefont {Itano}, \citenamefont {Wineland},\ and\ \citenamefont
  {Gould}}]{Monr95RSC}%
  \BibitemOpen
  \bibfield  {author} {\bibinfo {author} {\bibfnamefont {C.}~\bibnamefont
  {Monroe}}, \bibinfo {author} {\bibfnamefont {D.~M.}\ \bibnamefont {Meekhof}},
  \bibinfo {author} {\bibfnamefont {B.~E.}\ \bibnamefont {King}}, \bibinfo
  {author} {\bibfnamefont {S.~R.}\ \bibnamefont {Jefferts}}, \bibinfo {author}
  {\bibfnamefont {W.~M.}\ \bibnamefont {Itano}}, \bibinfo {author}
  {\bibfnamefont {D.~J.}\ \bibnamefont {Wineland}}, \ and\ \bibinfo {author}
  {\bibfnamefont {P.}~\bibnamefont {Gould}},\ }\href {\doibase
  10.1103/PhysRevLett.75.4011} {\bibfield  {journal} {\bibinfo  {journal}
  {Phys. Rev. Lett.}\ }\textbf {\bibinfo {volume} {75}},\ \bibinfo {pages}
  {4011} (\bibinfo {year} {1995})}\BibitemShut {NoStop}%
\bibitem [{\citenamefont {Hamann}\ \emph {et~al.}(1998)\citenamefont {Hamann},
  \citenamefont {Haycock}, \citenamefont {Klose}, \citenamefont {Pax},
  \citenamefont {Deutsch},\ and\ \citenamefont {Jessen}}]{Hamann98}%
  \BibitemOpen
  \bibfield  {author} {\bibinfo {author} {\bibfnamefont {S.~E.}\ \bibnamefont
  {Hamann}}, \bibinfo {author} {\bibfnamefont {D.~L.}\ \bibnamefont {Haycock}},
  \bibinfo {author} {\bibfnamefont {G.}~\bibnamefont {Klose}}, \bibinfo
  {author} {\bibfnamefont {P.~H.}\ \bibnamefont {Pax}}, \bibinfo {author}
  {\bibfnamefont {I.~H.}\ \bibnamefont {Deutsch}}, \ and\ \bibinfo {author}
  {\bibfnamefont {P.~S.}\ \bibnamefont {Jessen}},\ }\href@noop {} {\bibfield
  {journal} {\bibinfo  {journal} {Phys. Rev. Lett.}\ }\textbf {\bibinfo
  {volume} {80}},\ \bibinfo {pages} {4149} (\bibinfo {year}
  {1998})}\BibitemShut {NoStop}%
\bibitem [{\citenamefont {Vuleti\ifmmode~\acute{c}\else \'{c}\fi{}}\ \emph
  {et~al.}(1998)\citenamefont {Vuleti\ifmmode~\acute{c}\else \'{c}\fi{}},
  \citenamefont {Chin}, \citenamefont {Kerman},\ and\ \citenamefont
  {Chu}}]{Vuletic1998RSC}%
  \BibitemOpen
  \bibfield  {author} {\bibinfo {author} {\bibfnamefont {V.}~\bibnamefont
  {Vuleti\ifmmode~\acute{c}\else \'{c}\fi{}}}, \bibinfo {author} {\bibfnamefont
  {C.}~\bibnamefont {Chin}}, \bibinfo {author} {\bibfnamefont {A.~J.}\
  \bibnamefont {Kerman}}, \ and\ \bibinfo {author} {\bibfnamefont
  {S.}~\bibnamefont {Chu}},\ }\href {\doibase 10.1103/PhysRevLett.81.5768}
  {\bibfield  {journal} {\bibinfo  {journal} {Phys. Rev. Lett.}\ }\textbf
  {\bibinfo {volume} {81}},\ \bibinfo {pages} {5768} (\bibinfo {year}
  {1998})}\BibitemShut {NoStop}%
\bibitem [{\citenamefont {Kerman}\ \emph {et~al.}(2000)\citenamefont {Kerman},
  \citenamefont {Vuleti\ifmmode~\acute{c}\else \'{c}\fi{}}, \citenamefont
  {Chin},\ and\ \citenamefont {Chu}}]{Kerm00}%
  \BibitemOpen
  \bibfield  {author} {\bibinfo {author} {\bibfnamefont {A.~J.}\ \bibnamefont
  {Kerman}}, \bibinfo {author} {\bibfnamefont {V.}~\bibnamefont
  {Vuleti\ifmmode~\acute{c}\else \'{c}\fi{}}}, \bibinfo {author} {\bibfnamefont
  {C.}~\bibnamefont {Chin}}, \ and\ \bibinfo {author} {\bibfnamefont
  {S.}~\bibnamefont {Chu}},\ }\href@noop {} {\bibfield  {journal} {\bibinfo
  {journal} {Phys. Rev. Lett.}\ }\textbf {\bibinfo {volume} {84}},\ \bibinfo
  {pages} {439} (\bibinfo {year} {2000})}\BibitemShut {NoStop}%
\bibitem [{\citenamefont {Han}\ \emph {et~al.}(2000)\citenamefont {Han},
  \citenamefont {Wolf}, \citenamefont {Oliver}, \citenamefont {McCormick},
  \citenamefont {DePue},\ and\ \citenamefont {Weiss}}]{Han00}%
  \BibitemOpen
  \bibfield  {author} {\bibinfo {author} {\bibfnamefont {D.-J.}\ \bibnamefont
  {Han}}, \bibinfo {author} {\bibfnamefont {S.}~\bibnamefont {Wolf}}, \bibinfo
  {author} {\bibfnamefont {S.}~\bibnamefont {Oliver}}, \bibinfo {author}
  {\bibfnamefont {C.}~\bibnamefont {McCormick}}, \bibinfo {author}
  {\bibfnamefont {M.~T.}\ \bibnamefont {DePue}}, \ and\ \bibinfo {author}
  {\bibfnamefont {D.~S.}\ \bibnamefont {Weiss}},\ }\href {\doibase
  10.1103/PhysRevLett.85.724} {\bibfield  {journal} {\bibinfo  {journal} {Phys.
  Rev. Lett.}\ }\textbf {\bibinfo {volume} {85}},\ \bibinfo {pages} {724}
  (\bibinfo {year} {2000})}\BibitemShut {NoStop}%
\bibitem [{\citenamefont {Nelson}\ \emph {et~al.}(2007)\citenamefont {Nelson},
  \citenamefont {Li},\ and\ \citenamefont {Weiss}}]{nelson07}%
  \BibitemOpen
  \bibfield  {author} {\bibinfo {author} {\bibfnamefont {K.~D.}\ \bibnamefont
  {Nelson}}, \bibinfo {author} {\bibfnamefont {X.}~\bibnamefont {Li}}, \ and\
  \bibinfo {author} {\bibfnamefont {D.~S.}\ \bibnamefont {Weiss}},\ }\href@noop
  {} {\bibfield  {journal} {\bibinfo  {journal} {Nat. Phys.}\ }\textbf
  {\bibinfo {volume} {3}},\ \bibinfo {pages} {556} (\bibinfo {year}
  {2007})}\BibitemShut {NoStop}%
\bibitem [{\citenamefont {Kaufman}\ \emph {et~al.}(2012)\citenamefont
  {Kaufman}, \citenamefont {Lester},\ and\ \citenamefont
  {Regal}}]{kaufman2014}%
  \BibitemOpen
  \bibfield  {author} {\bibinfo {author} {\bibfnamefont {A.~M.}\ \bibnamefont
  {Kaufman}}, \bibinfo {author} {\bibfnamefont {B.~J.}\ \bibnamefont {Lester}},
  \ and\ \bibinfo {author} {\bibfnamefont {C.~A.}\ \bibnamefont {Regal}},\
  }\href {\doibase 10.1103/PhysRevX.2.041014} {\bibfield  {journal} {\bibinfo
  {journal} {Phys. Rev. X}\ }\textbf {\bibinfo {volume} {2}},\ \bibinfo {pages}
  {041014} (\bibinfo {year} {2012})}\BibitemShut {NoStop}%
\bibitem [{\citenamefont {Patil}\ \emph {et~al.}(2014)\citenamefont {Patil},
  \citenamefont {Chakram}, \citenamefont {Aycock},\ and\ \citenamefont
  {Vengalattore}}]{patil2014}%
  \BibitemOpen
  \bibfield  {author} {\bibinfo {author} {\bibfnamefont {Y.~S.}\ \bibnamefont
  {Patil}}, \bibinfo {author} {\bibfnamefont {S.}~\bibnamefont {Chakram}},
  \bibinfo {author} {\bibfnamefont {L.~M.}\ \bibnamefont {Aycock}}, \ and\
  \bibinfo {author} {\bibfnamefont {M.}~\bibnamefont {Vengalattore}},\ }\href
  {\doibase 10.1103/PhysRevA.90.033422} {\bibfield  {journal} {\bibinfo
  {journal} {Phys. Rev. A}\ }\textbf {\bibinfo {volume} {90}},\ \bibinfo
  {pages} {033422} (\bibinfo {year} {2014})}\BibitemShut {NoStop}%
\bibitem [{\citenamefont {Thompson}\ \emph {et~al.}(2013)\citenamefont
  {Thompson}, \citenamefont {Tiecke}, \citenamefont {Zibrov}, \citenamefont
  {Vuleti\ifmmode~\acute{c}\else \'{c}\fi{}},\ and\ \citenamefont
  {Lukin}}]{thompson14}%
  \BibitemOpen
  \bibfield  {author} {\bibinfo {author} {\bibfnamefont {J.~D.}\ \bibnamefont
  {Thompson}}, \bibinfo {author} {\bibfnamefont {T.~G.}\ \bibnamefont
  {Tiecke}}, \bibinfo {author} {\bibfnamefont {A.~S.}\ \bibnamefont {Zibrov}},
  \bibinfo {author} {\bibfnamefont {V.}~\bibnamefont
  {Vuleti\ifmmode~\acute{c}\else \'{c}\fi{}}}, \ and\ \bibinfo {author}
  {\bibfnamefont {M.~D.}\ \bibnamefont {Lukin}},\ }\href@noop {} {\bibfield
  {journal} {\bibinfo  {journal} {Phys. Rev. Lett.}\ }\textbf {\bibinfo
  {volume} {110}},\ \bibinfo {pages} {133001} (\bibinfo {year}
  {2013})}\BibitemShut {NoStop}%
\bibitem [{\citenamefont {Weiss}\ \emph {et~al.}(2004)\citenamefont {Weiss},
  \citenamefont {Vala}, \citenamefont {Thapliyal}, \citenamefont {Myrgren},
  \citenamefont {Vazirani},\ and\ \citenamefont {Whaley}}]{weiss2004}%
  \BibitemOpen
  \bibfield  {author} {\bibinfo {author} {\bibfnamefont {D.~S.}\ \bibnamefont
  {Weiss}}, \bibinfo {author} {\bibfnamefont {J.}~\bibnamefont {Vala}},
  \bibinfo {author} {\bibfnamefont {A.~V.}\ \bibnamefont {Thapliyal}}, \bibinfo
  {author} {\bibfnamefont {S.}~\bibnamefont {Myrgren}}, \bibinfo {author}
  {\bibfnamefont {U.}~\bibnamefont {Vazirani}}, \ and\ \bibinfo {author}
  {\bibfnamefont {K.~B.}\ \bibnamefont {Whaley}},\ }\href@noop {} {\bibfield
  {journal} {\bibinfo  {journal} {Phys. Rev. A}\ }\textbf {\bibinfo {volume}
  {70}},\ \bibinfo {pages} {040302} (\bibinfo {year} {2004})}\BibitemShut
  {NoStop}%
\bibitem [{\citenamefont {Park}\ \emph {et~al.}(2012)\citenamefont {Park},
  \citenamefont {Wu}, \citenamefont {Santiago}, \citenamefont {Tiecke},
  \citenamefont {Will}, \citenamefont {Ahmadi},\ and\ \citenamefont
  {Zwierlein}}]{Park:2012}%
  \BibitemOpen
  \bibfield  {author} {\bibinfo {author} {\bibfnamefont {J.~W.}\ \bibnamefont
  {Park}}, \bibinfo {author} {\bibfnamefont {C.-H.}\ \bibnamefont {Wu}},
  \bibinfo {author} {\bibfnamefont {I.}~\bibnamefont {Santiago}}, \bibinfo
  {author} {\bibfnamefont {T.~G.}\ \bibnamefont {Tiecke}}, \bibinfo {author}
  {\bibfnamefont {S.}~\bibnamefont {Will}}, \bibinfo {author} {\bibfnamefont
  {P.}~\bibnamefont {Ahmadi}}, \ and\ \bibinfo {author} {\bibfnamefont {M.~W.}\
  \bibnamefont {Zwierlein}},\ }\href@noop {} {\bibfield  {journal} {\bibinfo
  {journal} {Phys. Rev. A}\ }\textbf {\bibinfo {volume} {85}},\ \bibinfo
  {pages} {051602} (\bibinfo {year} {2012})}\BibitemShut {NoStop}%
\bibitem [{\citenamefont {Murmann}\ \emph {et~al.}(2015)\citenamefont
  {Murmann}, \citenamefont {Bergschneider}, \citenamefont {Klinkhamer},
  \citenamefont {Z\"urn}, \citenamefont {Lompe},\ and\ \citenamefont
  {Jochim}}]{murmann2015}%
  \BibitemOpen
  \bibfield  {author} {\bibinfo {author} {\bibfnamefont {S.}~\bibnamefont
  {Murmann}}, \bibinfo {author} {\bibfnamefont {A.}~\bibnamefont
  {Bergschneider}}, \bibinfo {author} {\bibfnamefont {V.}~\bibnamefont
  {Klinkhamer}}, \bibinfo {author} {\bibfnamefont {G.}~\bibnamefont {Z\"urn}},
  \bibinfo {author} {\bibfnamefont {T.}~\bibnamefont {Lompe}}, \ and\ \bibinfo
  {author} {\bibfnamefont {S.}~\bibnamefont {Jochim}},\ }\href@noop {}
  {\bibfield  {journal} {\bibinfo  {journal} {Phys. Rev. Lett.}\ }\textbf
  {\bibinfo {volume} {114}},\ \bibinfo {pages} {080402} (\bibinfo {year}
  {2015})}\BibitemShut {NoStop}%
\bibitem [{\citenamefont {Kaufman}\ \emph {et~al.}(2014)\citenamefont
  {Kaufman}, \citenamefont {Lester}, \citenamefont {Reynolds}, \citenamefont
  {Wall}, \citenamefont {Foss-Feig}, \citenamefont {Hazzard}, \citenamefont
  {Rey},\ and\ \citenamefont {Regal}}]{kaufman2014HOM}%
  \BibitemOpen
  \bibfield  {author} {\bibinfo {author} {\bibfnamefont {A.~M.}\ \bibnamefont
  {Kaufman}}, \bibinfo {author} {\bibfnamefont {B.~J.}\ \bibnamefont {Lester}},
  \bibinfo {author} {\bibfnamefont {C.~M.}\ \bibnamefont {Reynolds}}, \bibinfo
  {author} {\bibfnamefont {M.~L.}\ \bibnamefont {Wall}}, \bibinfo {author}
  {\bibfnamefont {M.}~\bibnamefont {Foss-Feig}}, \bibinfo {author}
  {\bibfnamefont {K.~R.~A.}\ \bibnamefont {Hazzard}}, \bibinfo {author}
  {\bibfnamefont {A.~M.}\ \bibnamefont {Rey}}, \ and\ \bibinfo {author}
  {\bibfnamefont {C.~A.}\ \bibnamefont {Regal}},\ }\href@noop {} {\bibfield
  {journal} {\bibinfo  {journal} {Science}\ }\textbf {\bibinfo {volume}
  {345}},\ \bibinfo {pages} {306} (\bibinfo {year} {2014})}\BibitemShut
  {NoStop}%
\bibitem [{\citenamefont {Wu}\ \emph {et~al.}(2012)\citenamefont {Wu},
  \citenamefont {Park}, \citenamefont {Ahmadi}, \citenamefont {Will},\ and\
  \citenamefont {Zwierlein}}]{wu2012NaK}%
  \BibitemOpen
  \bibfield  {author} {\bibinfo {author} {\bibfnamefont {C.-H.}\ \bibnamefont
  {Wu}}, \bibinfo {author} {\bibfnamefont {J.~W.}\ \bibnamefont {Park}},
  \bibinfo {author} {\bibfnamefont {P.}~\bibnamefont {Ahmadi}}, \bibinfo
  {author} {\bibfnamefont {S.}~\bibnamefont {Will}}, \ and\ \bibinfo {author}
  {\bibfnamefont {M.~W.}\ \bibnamefont {Zwierlein}},\ }\href@noop {} {\bibfield
   {journal} {\bibinfo  {journal} {Phys. Rev. Lett.}\ }\textbf {\bibinfo
  {volume} {109}},\ \bibinfo {pages} {085301} (\bibinfo {year}
  {2012})}\BibitemShut {NoStop}%
\end{thebibliography}

\end{document}